\documentclass[aps,preprint]{revtex4-1}

\usepackage{amsmath,amssymb,amsfonts,graphicx,hyperref,amsthm}

\newcommand{\beq}{\begin{equation}}
\newcommand{\eeq}{\end{equation}}
\newcommand{\bea}{\begin{eqnarray}}
\newcommand{\eea}{\end{eqnarray}}

\begin{document}

\title{Exact solutions and phenomenological constraints from massive scalars in a Gravity's Rainbow spacetime}

\author{V. B. Bezerra}
\email{valdir@fisica.ufpb.br}\address{Departamento de F\'{i}sica, Universidade Federal da Para\'{i}ba, Caixa Postal 5008, CEP 58051-970, Jo\~{a}o Pessoa, PB, Brazil}

\author{H. R. Christiansen}
\email{hugo.christiansen@ifce.edu.br}\address{Instituto Federal de Ci\^encia, Educa\c c\~ao e Tecnologia (IFCE),
 Departamento de F\'isica, Sobral 62040-730, Brazil.}

\author{M. S. Cunha }
\email{marcony.cunha@uece.br}\address{Grupo de F\'isica Te\'orica (GFT), Centro de Ci\^encias e Tecnologia, Universidade Estadual do Cear\'a, CEP 60714-903, Fortaleza, Cear\'a, Brazil.}

\author{C. R. Muniz}
\email{celio.muniz@uece.br}\address{Grupo de F\'isica Te\'orica (GFT), Universidade Estadual do Cear\'a, Faculdade de Educa\c c\~ao, Ci\^encias e Letras de Iguatu, Iguatu, Cear\'a, Brazil.}

\begin{abstract}

We obtain the exact (confluent Heun) solutions to the massive scalar field in a Gravity's Rainbow Schwarzschild metric. With these solutions at hand, we study the Hawking radiation resulting from the tunneling rate through the event horizon. We show that the emission spectrum obeys non-extensive statistics and is halted when a certain mass remnant is reached.
Next, we infer constraints on the rainbow parameters from recent LHC particle physics experiments and Hubble STIS astrophysics measurements. Finally, we study the low frequency limit in order to find the modified energy spectrum around the source.

\vspace{0.75cm}
\end{abstract}

\maketitle

\section{Introduction}

The meaning of the theory of General Relativity (GR) has been brilliantly resumed
by J. A. Wheeler in a simple and celebrated sentence:
``Matter tells spacetime how to curve, and spacetime tells matter how to move''.
This aphorism emphasizes the {\it fulcrum} of the Einstein's theory, according to which the gravitational field
reveals itself as a curvature of spacetime \cite{Wheeler}.
This picture works very well for a massive body curving the spacetime where a (not very high-energy)
particle is moving in. However, at ultra high-energies the theory needs serious revision for the concept
of spacetime coordinates itself has to be reconsidered.

Indeed, the formal necessity of a tiny-scale (so-called UV) modification of GR arises
from the non-renormalizability of a quantum field theory of gravity.
This can be seen from the high-energy divergent diagrams in Feynman
loop-expansion \cite{stelle}.
A way to deal with this issue is to insert higher-order derivative terms in the Lagrangian which modify
the UV graviton propagator \cite{stelle}. But the problem is that higher-order time derivatives in the equations
of motion lead to ghosts. In order to keep the good and remove the bad,
P. Horava introduced a asymmetric Lifshitz scaling between space and time \cite{Horava}
such that higher order spatial-derivatives are not accompanied by higher order time- ones in the UV regime.
This particular Lorentz symmetry violation
allows power-counting renormalizability while avoids ghosts and GR arises as an infrared fixed point.

An alternative approach is to modify the metric instead of reshaping the action.
This so-called \emph{Rainbow Gravity} scenario \cite{MS2004}
radically changes the GR framework since, according to it,
the UV deformed metric introduces the asymmetry between space and time but now
it is done through the energy of the probe.
This correction has been shown to fix
one loop divergences, avoiding the need of a renormalization scheme \cite{garattini}. This is a great advantage
which makes the rainbow proposal a source of constant investigation, e.g.\cite{2017}
(a connection with the Horava-Lifshitz gravity can be found in \cite{garattini2}).

Rainbow gravity can be obtained as a generalization to curved spacetimes
of the so-called Doubly Special Relativity (DSR) \cite{MSmolin1,MSmolin2,Camelia}.
In DSR the transformation laws in energy-momentum space are nonlinear. The dual space, $(x,t)$, is thus endowed
with a nontrivial quadratic invariant, namely, an energy-dependent metric tensor.
It means that if a given observer measures a particle (or wave)
with energy $E$, then he concludes that this
probe feels a metric $\mathrm{g}_{ab}(E)$. But, on the other hand, a different observer will measure
instead $E'\neq E$, and will therefore assign to the particle's
motion different metric tensor elements $\mathrm{g}_{ab}(E')$.
This argument, valid in flat spacetime, carries over locally to curved spacetimes using the equivalence principle.
As a consequence, the invariant norm is no longer bilinear and leads to modified dispersion relations.

In DSR the laws of energy-momentum conservation still
hold in all inertial frames but then again they are non-linear. One important reason for
the DSR proposal is that the energy scale which establishes the boundary between the
quantum and classical characterization of spacetime,  $E_P$, can be assumed as an invariant in the
sense that all inertial observers agree on whether a particle has more, or less, than
this energy. Interestingly, this works out an otherwise inconvenience; namely, that the threshold between the
quantum and the classical description can depend on the speed of the observer \cite{MS2004}.

The first accomplishment in DSR \cite{Amati} implies a deformed Lorentz symmetry such
that the standard energy-momentum relations in flat spacetime are modified
by Planck scale corrections of the form
\begin{equation}\label{MDR}
E^2\,f^2(\omega/\omega_P)-(pc)^2\,g^2(\omega/\omega_P)=m^2 c^4,
\end{equation}
where $m$ is the rest frame mass of the particle with energy-frequency
$\omega$ as seen by an inertial observer, and $\omega_P$ is the Planck
energy-frequency $E_P=\hbar \omega_P$. Global Lorentz invariance is in fact
an accidental symmetry related to a particular solution of General Relativity.
Thus, whether it is broken or modified it is only a symmetry emerging at low energies
from a quantum theory of gravity and is just approximate.

It is generally accepted that at sufficiently high energies the geometry of spacetime should be described by
a quantum theory in which General Relativity is replaced by a
quantum mechanical description of the spacetime coordinates.
It is also believed that the Planck energy $E_P = \sqrt{\hbar c^5/G}$ establishes the threshold
that separates the classical description from the quantum description of gravity.
Above the Planck scale a continuous spacetime manifold loses consistency for quantum effects become
uncontrollable and a metric approach becomes impracticable. Thus, rainbow gravity is concerned with
the effects on the propagation of particles with energies
below $E_P$ but whose wavelengths are much shorter than the local radius of curvature of spacetime.
Of course, to be consistent with the standard theory, the functions $f(\omega/\omega_P)$ and
 $g(\omega/\omega_P)$ which appear in Eq. (\ref{MDR}) must tend to unity near $E_P$.
In this context, a generalized uncertainty principle (GUP) is often introduced in order to account for this
fuzzy microscopic structure of spacetime and to avoid the singularities
of the general relativity \cite{Maggiore1,Maggiore2}.

In this work we will adopt a semiclassical approach inspired in loop quantum gravity \cite{Smolin0}
to study an uncharged scalar field placed in a spherically symmetric spacetime characterized by
the aforementioned MDR functions defined as
\beq f(\omega/\omega_P)=1, \ \ \ \ \ \ g(\omega/\omega_P)=\sqrt{1-\xi(\omega/\omega_P)^s},\eeq
where $\xi>0$, and $s$ is a positive integer of order one. We will obtain the exact wave solutions to the
scalar field and later on analyze them near the event horizon in order to compute the Hawking radiation via
quantum tunneling through this frontier. Constraints on these rainbow parameters will be obtained by considering
particle physics experiments related to negative results regarding the creation of microscopic black holes
in the LHC as well as galactic measurements of the fine-structure constant made
by the Hubble Space Telescope (STIS) from a white dwarf spectrum.

The paper is organized as follows: In section II we obtain the exact solutions of a massive scalar
field in the rainbow Schwarzschild metric, then study the Hawking radiation, and thereafter calculate
the energy eigenspectrum and infer new constraints on the rainbow parameters.
Finally, in section III, we draw the conclusions.

\section{Massive Scalar Field in Rainbow Schwarzschild Spacetime}

{Although it has been a long-studied subject, the massive scalar field in a Schwarzschild spacetime
(see \cite{Frolov} and references therein) lacked of an exact solution until recently \cite{Fiziev3}.
Now, it is known that its whole space spectrum is formally given in terms of Heun's functions \cite{Ronveaux}
combined with elementary functions. Here we will employ an analytic approach in order to solve
such a problem, this time considering a gravity's rainbow metric.}

\subsection{Solutions}
Our task is solving the rainbow gravity covariant Klein-Gordon equation of massive scalars
 minimally coupled to the Schwarzschild gravitational field
\begin{equation}
\left[\frac{1}{\sqrt{-\mathrm{g}}}\partial_{\mu}\left(\mathrm{g}^{\mu\nu}\sqrt{-\mathrm{g}}
\partial_{\nu}\right)+m^{2}\right]\Psi=0\ ,
\label{eq:Klein-Gordon_cova}
\end{equation}
(where natural units $c \equiv \hbar \equiv 1$ are used).

The gravitational background generated by a static uncharged compact object is given
by the Schwarzschild metric now depending on the rainbow functions $f(\omega/\omega_P)$ and $g(\omega/\omega_P)$.
In spherical coordinates the square line-element invariant reads  \cite{JMathPhys.8.265}
\begin{equation}
ds^{2}=f^{-2}(\omega/\omega_P)h(r)dt^{2}-g^{-2}(\omega/\omega_P)[h(r)^{-1}dr^{2}-r^{2}d\Omega^2],
\label{eq:metrica_Kerr-Newman}
\end{equation}
where $h(r)=\left(1-\frac{r_s}{r}\right)$, $r_s=2MG$ is the Schwarzschild radius,
 $d\Omega^2=d\theta^{2}+\sin^{2}\theta\ d\phi^{2}$, $G=G(0)$
is the Newton's universal gravitational constant and $M$ is the mass of the source.
By symmetry arguments we assume that solutions of Eq.~(\ref{eq:Klein-Gordon_cova}) can be factored as follows
\begin{equation}
\Psi(\mathbf{r},t)=R(r)Y_{l}^{m_l}(\theta,\phi)\mbox{e}^{-i\omega t},
\label{eq:separacao_variaveis}
\end{equation}
where $Y_{l}^{m_l}(\theta,\phi)$ are the spherical harmonic
 functions. Inserting Eq. (\ref{eq:separacao_variaveis})
and the metric given by Eq.~(\ref{eq:metrica_Kerr-Newman}) into (\ref{eq:Klein-Gordon_cova}),
we obtain the following radial equation
\begin{equation}
\frac{d}{dr}\left[r(r-2GM)\frac{dR}{dr}\right]
+\left(\frac{r^{3}\tilde{\omega}^{2}}{r-2GM}-\tilde{m}^{2}r^{2}-\lambda_{lm_l}\right)R=0,
\label{eq:mov_radial_1}
\end{equation}
where $\lambda_{lm_l}=l(l+1)$ and
\begin{eqnarray}\label{tildes}
 \tilde{\omega}&=&\frac{\omega}{g(\omega/\omega_P)}, \\ \nonumber
 \tilde{m}&=&\frac{m}{g(\omega/\omega_P)}.
 \end{eqnarray}

The expression given by Eq. (\ref{eq:mov_radial_1}) has singularities
at $r=(a_{1},a_{2})=(0, 2GM)$ and $\infty$,
and can be transformed into a Heun equation by using
\begin{equation}
x=\frac{r-a_{1}}{a_{2}-a_{1}}=\frac{r-2GM}{2GM}\ .
\label{eq:homog_subs_radial}
\end{equation}
Let us introduce the function $Z(x)$ such that
\beq R(x)=Z(x)[x(x-1)]^{-1/2},\eeq
and set henceforth $G=1$.
 Then, differential Eq.~(\ref{eq:mov_radial_1}) transforms into
\begin{equation}
\frac{d^{2}Z}{dx^{2}}+\left[A_{1}+\frac{A_{2}}{x}+
\frac{A_{3}}{x-1}+\frac{A_{4}}{x^{2}}+\frac{A_{5}}{(x-1)^{2}}\right]Z=0 ,
\label{eq:mov_radial_x_heun}
\end{equation}
where the coeficients $A_{1}$, $A_{2}$, $A_{3}$, $A_{4}$, and $A_{5}$ are given by
\begin{equation}
A_{1}=-4 M^2 \left(\tilde{m}^2-\tilde{\omega}^2\right)\ ;
\label{eq:A1_mov_radial_x_normal}
\end{equation}

\begin{equation}
A_{2}=\frac{1}{2}+\lambda_{lm_l} +4 M^2 \left(\tilde{m}^2-2 \tilde{\omega}^2\right)\ ;
\label{eq:A2_mov_radial_x_normal}
\end{equation}

\begin{equation}
A_{3}=-\frac{1}{2}-\lambda_{lm_l}\ ;
\label{eq:A3_mov_radial_x_normal}
\end{equation}

\begin{equation}
A_{4}=\frac{1}{4}+4 M^2 \tilde{\omega}^2\ ;
\label{eq:A4_mov_radial_x_normal}
\end{equation}

\begin{equation}
A_{5}=\frac{1}{4}\ .
\label{eq:A5_mov_radial_x_normal}
\end{equation}

The general solution to Eq.~(\ref{eq:mov_radial_x_heun}) over the entire
range $0<x\leq \infty$ is given by \cite{Horacio2}
\begin{eqnarray}
R(x) & = &C_{1} \mbox{e}^{\frac{1}{2}\alpha x}x^{\frac{1}{2}\beta}
\mbox{HeunC}(\alpha,\beta,\gamma,\delta,\eta;x)\nonumber\\
&+&C_{2}\mbox{e}^{\frac{1}{2}\alpha x}\ x^{-\frac{1}{2}\beta}
\mbox{HeunC}(\alpha,-\beta,\gamma,\delta,\eta;x)\ ,
\label{eq:solucao_geral_radial_Kerr-Newman_gauge}
\end{eqnarray}
where $C_{1}$ and $C_{2}$ are constants, and the parameters $\alpha$, $\beta$, $\gamma$, $\delta$,
and $\eta$ explicitly written in terms of the rainbow's function are given by:

\begin{subequations}

\begin{equation}
\alpha=-4M\sqrt{\frac{m^{2}-\omega^{2}}{1-\xi(\omega/\omega_P)^s}} ;
\label{eq:alpha_radial_HeunC_Kerr-Newman}
\end{equation}

\begin{equation}
\beta=\frac{i4M\omega}{\sqrt{1-\xi(\omega/\omega_P)^s}}\ ;
\label{eq:beta_radial_HeunC_Kerr-Newman}
\end{equation}

\begin{equation}
\gamma=0\ ;
\label{eq:gamma_radial_HeunC_Kerr-Newman}
\end{equation}

\begin{equation}
\delta=\frac{4M^{2}\left(m^{2}-2\omega^{2}\right)}{1-\xi(\omega/\omega_P)^s};
\label{eq:delta_radial_HeunC_Kerr-Newman}
\end{equation}

\begin{equation}
\eta=-l(l+1)-\frac{4M^{2}\left(m^{2}-2\omega^{2}\right)}{1-\xi(\omega/\omega_P)^s}.
\label{eq:eta_radial_HeunC_Kerr-Newman}
\end{equation}

\end{subequations}

This is the sum of two linearly independent solutions of the confluent Heun differential
equation provided $\beta$ is not an integer \cite{Ronveaux}.

{It is also worth mentioning some pioneering work done in this direction \cite{japan3},
where analytic solutions to the (massless) Regge-Wheeler and Teukolsky equations are found as a series of
hypergeometric and Coulomb  wavefunctions with different regions of convergence. This has
been used in \cite{binidamour} to compute the post-Minkowskian expansion of Regge-Wheeler-Zerilli 
black hole perturbation theory to calculate
a fourth order post-Newtonian approximation of the main radial
potential describing the gravitational interaction of two bodies. For a massive scalar particle, 
the effects of the self-force upon the orbits of a Schwarzschild black hole have been computed in \cite{detweiler}.
}

\subsection{Uncompleted Hawking Radiation}

The exterior outgoing wave solutions at the event horizon of a Schwarzschild black hole
are obtained by taking $x\rightarrow 0^+$ in Eq. (\ref{eq:solucao_geral_radial_Kerr-Newman_gauge})
for positive frequencies.
If we also consider the temporal part of the wave function, the result is
\begin{equation}
\Psi_{out}(r\simeq 2M^+)\simeq C_1e^{-i\omega t}(r-2M)^{2iM\tilde\omega}.
\end{equation}
Let us now examine the variable
\beq r_*=g^{-1}\, (\omega/\omega_P)\left[r+2M \log{\left|\frac{r-2M}{2M}\right|}\right]\eeq
inspired in the conventional $(\xi=0)$ tortoise coordinate (which approaches $-\infty$ when $r\rightarrow r_s$),
appropriate to analyze perturbations in the spherically symmetric gravitational field \cite{Frolov}.
Considering Eddington-Finkelstein coordinates, we define $v = t+ r_*$ yielding
\begin{equation}\label{OSTC}
\Psi_{out}(r\simeq 2M^+)=C_1e^{-i(\omega v-\tilde{\omega} r)}(2M)^{-2iM\tilde{\omega}}(r-2M)^{4iM\tilde{\omega}}.
\end{equation}
Following \cite{Damour} we now consider the analytic extension of the solution
to the interior region ($r\simeq 2M^-$) by means of a rotation in the complex plane
 $(r-2M)\rightarrow (2M-r)e^{-i\pi}$ in Eq.(\ref{OSTC}). Thus, one has
\begin{equation}
\Psi_{out}(r\simeq 2M^-)=C_1e^{-i(\omega v-\tilde{\omega} r)}(2M)^{-2iM\tilde{\omega}}
(2M-r)^{4 i M\tilde{\omega}}e^{4 \pi M\tilde{\omega}}.
\end{equation}
With these two expressions we can calculate the transmission coefficient through the horizon,
namely the tunneling rate, defined as
\begin{equation}\label{rate-tunnel}
\Gamma_{ev}=\left|\frac{\Psi_{out}(r\simeq 2M^+)}{\Psi_{out}(r\simeq 2M^-)}\right|^2=
\exp{\left[-\frac{8\pi M \omega}{g(\omega/\omega_P)}\right]}.
\end{equation}

It is worth noting that rainbow gravity black holes never evaporate completely.
Unlike the ordinary Schwarzschild black hole, the rainbow one halts outwards tunneling
when it reaches a certain nonzero minimal value which equals the critical mass (i.e.
the mass which avoids the BH temperature turning imaginary) \cite{remnant}.
This remnant mass can be calculated by making $\Gamma_{ev}=0$ in Eq. \ref{rate-tunnel}
assuming a generalized position-momentum uncertainty principle.

This GUP can be motivated on general grounds by the intuition
that the solution of the quantum gravity problem would need an absolute planckian limit
of the size of the collision region \cite{Amelino_IJMP, Amelino_CQG}.
So far, it is consensual  both in String Theory and  Loop Quantum Gravity
that a GUP compatible with a (leading order correction) logarithmic-area growth
of BH entropy should be of the the form
\beq \delta x \geq \frac1{\delta p}+ \lambda L_P^2 \delta p +O(L_P^3\delta p^2)\eeq
where the coefficient $\lambda$ should take a value of roughly the ratio between the square of the string
length and the square of the Planck length $L_P$. While in nonrelativistic quantum mechanics a particle of
any energy can always be sharply localized (at the price of losing any information on the conjugate momentum),
within quantum field theory it can only happen in the infinite-energy limit. However,
at the quantum gravity level the intuition is that such a sharp localization should disappear
 and uncertainty  could be recoded in a relation of the type
\beq E\geq \frac1{\delta x} [1-\Delta(L_P,\delta x)]\label{EGUP}\eeq
where $\Delta(L_P,\delta x)$ should be such that $E\rightarrow \infty$ at some nonzero $\delta x$.
According to the usual argument of quantum mechanics, when the position of a particle of mass $M$ (at rest)
is being measured by a procedure involving a collision with a photon of momentum $p_\gamma$,
we have $\delta p_\gamma\geq 1/\delta x$ where $\delta p_\gamma$ is the photon momentum uncertainty and $\delta x$
is the position uncertainty of the $M$ particle. Using the special relativity Heisenberg's uncertainty principle,
it also means that $\delta E_\gamma\geq 1/\delta x$ which yields naturally  $M\geq 1/\delta x$ since we need
$\delta E_\gamma\leq M$ in order not to disturb completely the system being measured.
Applying a boost the relation results in $E\geq 1/\delta x$, which on a rainbow gravity
basis carries into $E\geq [1+\sum a_k (L_P/\delta x)^{k}]/\delta x$, as stated in Eq. (\ref{EGUP}).

In our context, assuming that the test particle is a massless scalar localized within the BH event horizon  $r_s=2M$,
its frequency uncertainty results $\delta\omega\geq (2M)^{-1}$ which at leading order implies then again
$\omega\geq (2M)^{-1}$  \cite{Ali}.
The remnant mass of the BH, $M_r$, can be therefore obtained by making $g(\omega/\omega_P)=0$ in Eq. (\ref{rate-tunnel}),
i.e. considering null evaporation rate for the minimal $\omega$ value.
After some straightforward calculation, we obtain the following expression
for $M=M_r$ as a function of the rainbow parameters
\begin{equation}\label{critical-mass}
M_r=\frac{\xi^{1/s}}{2}M_P,
\end{equation}
where $M_P$ is the Planck mass.

Should we know the phenomenological mass value of the remnant, we could calculate a lower bound to the rainbow parameters.
From recent negative results regarding microscopic black holes in the CMS experiment at CERN's Large Hadron Collider
we can so far exclude BH masses below 6.2 TeV \cite{Sirunyan}. This allows establishing a constraint in the parameters given by
\begin{equation}\label{constraint-LHC}
6.2 \mathrm{TeV}<\frac{\xi^{1/s}}{2}2\times 10^{16} \mathrm{TeV}\Rightarrow \xi^{1/s}>6.2\times 10^{-16},
\end{equation}
which for a conservative (simple) assumption of $s=1$ and $s=2$, result in
 $\xi>6.2 \times 10^{-16}$ and $\xi>3.8\times 10^{-31}$, respectively.
Other constraints on this parameter calculated from data obtained in the ATLAS experiment
were pointed out in \cite{Ali2}, including those related to extra-dimensional considerations.

{Let us now focus on the \emph{grey-body} spectrum emitted from the rainbow black hole. Its
distribution function, or occupation number $n_{\omega}$, is given by \cite{Horacio3}
\begin{equation}\label{spectrum}
n_{\omega}=\frac{\Gamma_{ev}}{1-\Gamma_{ev}}=\frac{1}{e^{^q(\omega)8\pi M\omega}-1},
\end{equation}
where $q(\omega)=g^{-1}(\omega/\omega_P)$ happens to be the Tsallis parameter associated with an
incomplete non-extensive entropy
(see \cite{Wang} and references therein) and ${8\pi M}$ is the inverse of the usual Hawking temperature
associated with a blackbody (Planck) spectrum, ${T}^{-1}_{H}$. }

{ Nonextensivity has already been found in astrophysical contexts
associated with the dynamics of systems under long range interactions at long and short distances
\cite{tsallis, japan, chaos, cbpf, PLB2016}.
The nonextensive statistical parameters have been shown to have important physical
meaning in dictating the final mass of the formed
black hole through the scaling laws found in the asymptotic
regimes of strong and weak rates of mass loss, respectively \cite{cbpf}. }

{At ultra high energies, there is an alternative to the GUP above discussed.
Since gravitational back reaction dodge testing spacetime, its description as a smooth manifold appears as
a practical mathematical hypothesis. It comes then natural to soften this assumption
and conceive a more general \textit{noncommutative} discretized spacetime endowed with uncertainty
relations  among the spacetime coordinates themselves. Thus, noncommutative geometry gets into matter also.
A possible connection between results in these two scenarios is found in \cite{chaos}
by means of a relation between the parameters of the corresponding theories through nonextensive thermodynamics
outcomes.}

{Note that differently from previous work on the subject, the expression found above, Eq. (\ref{spectrum}),
shows a significant deviation from the usual spectrum, particularly because the parameter now depends
on the particle's energy.}
Indeed, from Eq. (\ref{spectrum}) we can see that the Hawking temperature in rainbow gravity can be defined by
\begin{equation}\label{T}
\tilde{T}_{H}=\frac{1}{8\pi M\,\,q(\omega)}=\frac{1}{8\pi M}\sqrt{1-\xi\left(\frac{\omega}{\omega_P}\right)^s},
\end{equation}
which as expected returns the standard ${T}_{H}$ value for $\xi= 0$.
In Fig.1 below we show (a magnification of) the dependence of the emitted energy density per frequency unit, $\rho(\omega)=\omega^3n_\omega$ for some values of $\xi$.
Notice that the resulting UV energy density (above 0.2) gets lower as the rainbow parameter grows.
It might be interpreted as a shift resulting from the fact that the particle's energy is being partially spent
in deforming the spacetime otherwise unaffected, as in ordinary GR. On the other hand,
the spectral emissivity grows in the middle region, near the maximum.
As expected, the curves are practically indistinguishable at low non-planckian energies.

\begin{figure}[!h]
\centering
\includegraphics[scale=0.8]{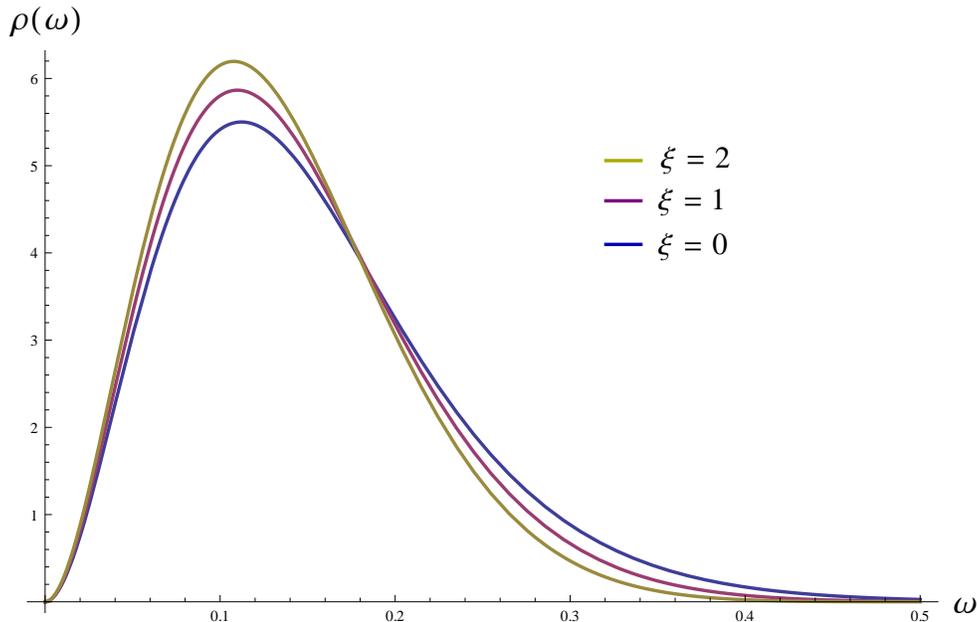}
\caption{Energy density $\rho(\omega)=\omega^3n_\omega$
emitted by a rainbow black hole for growing values of the rainbow
parameter, $\xi$, as a function of the frequency. We set $s=2$ and $T=0.1$ in Planck units.}
\end{figure}

\subsection{Variable Fine Structure Constant}

Now, our aim is setting an upper bound on  $\xi$ from current astrophysical measurements.
Making explicit the linear dependence of $\tilde{T}_{H}$ with $\hbar$ in eq.(\ref{T}),
we can ascribe all the rainbow dependence to a modified Planck constant, and define
\begin{equation}\label{effhbar}
\tilde{\hbar}=\hbar g(\omega/\omega_P)=\hbar \sqrt{1-\xi\left(\frac{\omega}{\omega_P}\right)^s}.
\end{equation}
Now, since the fine structure constant ${\alpha}$ is inversely proportional to $\hbar$ we can
define a modified fine structure $\tilde{\alpha}$
\begin{equation}\label{alphaeff}
\tilde{\alpha}=\frac{\alpha}{\sqrt{1-\xi\left(\frac{\omega}{\omega_P}\right)^s}}\approx \alpha\left[1+\frac{\xi}{2}\left(\frac{\omega}{\omega_P}\right)^s\right]
\end{equation}
to take place in strong gravitational scenarios.

Hence, considering the most recent data on the relative fine structure constant
in the gravitational field of a white dwarf,
one verifies that  $\Delta\alpha/\alpha\approx 10^{-5}$ \cite{Barrow} as registered by the
Hubble Space Telescope STIS from the absorbtion spectra of metal lines of G191-B2B.
From this we can get a constraint on $\xi$ given by
\begin{equation}\label{xiconstraint}
10^{-5}\gtrsim \frac{\xi}{2}\left(\frac{\omega}{\omega_P}\right)^s\Rightarrow
\xi\lesssim 2\times10^{-5} \left(\frac{\omega_P}{\omega}\right)^s.
\end{equation}
For a surface temperature of 60,000$-$70,000 K \cite{Rauch}, this yields $\xi\lesssim 4\times 10^{22}$ for $s=1$.
This upper bound gets of course higher for higher values of $s$ (see \cite{Steinhardt} for a discussion of some theoretical models).
A comprehensive investigation in Ref. \cite{MNRAS2012} suggests that a spatial variation of $|\Delta\alpha/\alpha|$ is in
general compatible with $10^{-5}$ which puts the same upper bound on $\xi$.
Finally, in the domain of atomic physics the authors of \cite{rubidio} obtain a bound to the uncertainty
in the gravitational gradient in an experiment
involving the recoil velocity of $^{87}Rb$ atoms in a vertical lattice. In this case it yields $\xi\lesssim 2.8\times 10^{23}$ for $s=1$.
Constraints on $s$ are also exhibited in Table I below, for these scenarios.


\subsection{Gravitational Quantum Energy Levels}

In this subsection, we will determine the energy eigenvalues of the massive scalar in a strong gravitational field
by setting boundary conditions at the asymptotic region. In order to annul the general
solution at infinity we impose a necessary condition in Eq. (\ref{eq:solucao_geral_radial_Kerr-Newman_gauge})
which guarantees $R(x)$ to be a finite polynomial.

We start considering the confluent solution in the disk $|z|<1$ defined by the series expansion
\bea
\text{HeunC}(\alpha,\beta,\gamma,\delta,\eta,z)=\sum_{n=0}^\infty v_{n}(\alpha,\beta,\gamma,\delta,\eta)z^n,
\label{HeunC}
\eea
together with the condition $\text{HeunC}(\alpha,\beta,\gamma,\delta,\eta,0)=1$.
The coefficients $v_{n}(\alpha,\beta,\gamma,\delta,\eta)$ are determined by a three-term recurrence relation
\beq
A_{n}v_{n}=B_{n}v_{n-1}+C_{n}v_{n-2}
\label{recurrence}
\eeq
with initial conditions $v_{-1}=0,\,\,v_{0}=1$ \cite{Fiziev}. Here
\bea
\hskip -.truecm A_{n}&=&1+{\frac{\beta}{n}} \\
\hskip -.truecm B_{n}&=&1+{\frac{-\alpha+\beta+\gamma-1}{n}}+
              {\frac{\eta-(-\alpha+\beta+\gamma)/2-\alpha\beta/2+\beta\gamma/2}{n^2}} \\
\hskip -.truecm C_{n}&=&{\frac{\alpha}{n^2}}\left({\frac \delta \alpha}+{\frac {\beta+\gamma}{2}}+n-1\right).
\label{rec_coeff}
\eea
Thus, in order to have a polynomial confluent Heun function
(\ref{eq:solucao_geral_radial_Kerr-Newman_gauge}),
we must impose the so called $\delta_N$ and $\Delta_{N+1}$
conditions
\begin{eqnarray}
\frac{\delta}{\alpha}+\frac{\beta+\gamma}{2}+1&=&-N \label{eq:cond_polin_1}\\
\Delta_{N+1}&=&0
\label{eq:cond_polin_2}
\end{eqnarray}
where $N$ is a non-negative integer \cite{Ronveaux}. For further details see also \cite{Fiziev}.
From Eq. \ref{eq:cond_polin_1}, we obtain the following expression for the energy levels
\begin{equation}
n+g^{-1}(\omega/\omega_P)\left[2iM\omega-\frac{M\left(2\omega^{2}-m^{2}\right)}{\sqrt{m^{2}-\omega^{2}}}\right]=0,
\label{eq:energy_levels}
\end{equation}
where $n=N+1$.

Let us now consider the low energy regime $\omega M \ll 1$, in which the particle
probe is not absorbed by the black hole.
According to Eq. (\ref{rate-tunnel}), in the limit there will be no tunneling into the event horizon
of the rainbow black hole since
\begin{equation}
\lim_{\omega M\rightarrow 0} e^{-8\pi M\tilde{\omega}}=1
\end{equation}
and therefore $\Gamma_{abs}=1-\Gamma_{ev}\rightarrow 0.$

Stationary bound-state solutions are formed by waves that propagate in opposite directions, with $\omega\in \Re$.
In the present case, these are sums of  outward matter waves coming from the event horizon
superposed with inward matter waves  moving towards the horizon,
thereafter tunneling in through the Regge-Wheeler barrier \cite{Fiziev4}.
Interestingly,  the condition of no waves coming out from (nor going into) the horizon
introduces complex valued frequencies which correspond to quasi-bound states \cite{BhabaniGrain}.

Rewriting Eq. \ref{eq:energy_levels}
for $\omega M \ll 1$, we obtain
\begin{equation}\label{eigenfrequencies-equation}
\xi\omega^{s+2}-\xi m^2\omega^s-\omega^2+2m\omega^{(0)}_n+m^2=0,
\end{equation}
where $\omega_n^{(0)}=-m^3M^2/2n^2$, with $n=1,2,3...$, are the gravitational Bohr levels
(recall that in the present units, $\omega_P=1$).

Interestingly, for  $s=1$ equation (\ref{eigenfrequencies-equation})
 has two complex omega solutions and a real one but
it diverges as $\xi$ approaches zero which we also disregard.
For $s=2$ there is only one relevant solution and it is given by
\begin{equation}\label{eigenfrequencies-solution}
\omega_n=\sqrt{\frac{m^2}{2}+\frac{1}{2\xi }-\frac{\sqrt{m^4 \xi ^2-2 m^2 \xi -8 m \xi\omega_n^{(0)}+1}}{2\xi }}.
\end{equation}
For $\xi\rightarrow 0$ the corresponding bound state energies coincide with the energy spectrum given
in \cite{Barranco, CelioEPL}.
At first order in $\xi$,  the gravitational Bohr levels are given by
\begin{equation}\label{omega}
\omega_n^2\approx m^2 + 2m \omega_n^{(0)}+\; \xi (2m^3\omega_n^{(0)}+4m^2\omega_n^{(0)^2} )+ \dots .
\end{equation}
 %
\section{Closing remarks}

In this paper, we have presented the analytic solution to the Klein-Gordon equation of a massive scalar in
the gravity's rainbow Schwarzschild spacetime.
Analyzing both the exterior and interior outgoing wavefunctions
at the event horizon of a black hole we calculated the tunneling rate of test particles through
this boundary. We have demonstrated that black hole evaporation is incomplete and stops at a remnant mass
compatible with a generalized uncertainty principle.
{Next, we have evinced that the scalar emission spectrum (emitted particle occupation number) is associated
with an incomplete non-extensive statistics where the Tsallis parameter is the rainbow function $g^{-1}(E/E_P)$.
Remarkably, non-extensivity is here not merely realized through some constant $q\neq 1$, as in the
available literature on BH,
but by means of a function of the particle's energy and the rainbow parameters.
The Hawking temperature is thereby modified as exhibited in Eq. \ref{T}.
Using this connection, a lower constraint on $\xi$ was obtained by means of recent LHC negative
results related to microscopic black holes.}

Thereafter, we calculated the gravity deformed fine structure constant $\tilde{\alpha}$ in terms of the
rainbow function $g(\omega/\omega_P)$. Astrophysical measurements of $\Delta\alpha/\alpha$
allowed us setting an upper constraint on the rainbow parameters.
Finally, we computed the stationary eigenenergy modes of the massive scalar field
in the low energy regime $M \omega \ll 1$ in which the particle probe does not tunnel through the horizon.
Whence, we obtained the rainbow gravity corrected analog of the Bohr levels for the hydrogen atom. We have
solved the corresponding equation for $s=1$ and $s=2$ and found that no meaningful solution exists in the first case.
The second case has just one physically relevant solution which converges to the ordinary levels as $\xi\rightarrow 0$
and whose rainbow first order correction is given in Eq. (\ref{omega}).

The table below resumes our results on the constraints to the gravity's rainbow parameters,
as compared with others registered in the literature and collected from entirely different experiments.

\begin{table}[h]
\caption{Rainbow's gravity parameter constraints}
\centering
\begin{tabular}{|l|c|c|c|}\hline
{\bf Experiment} & {\bf On $\xi$ ($s=1$)} & {\bf On $\xi$ ($s=2$) } & {\bf On $s$ ($\xi= 0.5$)} \\ \hline
Black holes in LHC & $> 6\times 10^{-16}$ & $> 4\times 10^{-31}$ & $< 0.02$\\ \hline
Variable $\alpha$ (white dwarfs) & $< 4\times10^{22}$ & $< 10^{43}$ & $> 0.16$ \\ \hline
Photon time delay \cite{Ali3}& $<  10^{20} $ & $-$ & $-$  \\ \hline
Weak equivalence principle (rotating torsion balance) \cite{Ali3} & $< 10^5$ &$-$& $-$  \\ \hline
\end{tabular}
\end{table}

%
\section*{Acknowledgements}

The authors would like to thank Conselho Nacional de Desenvolvimento Cient\'{i}fico e Tecnol\'{o}gico (CNPq) for financial support.


\end{document}